\documentclass[12pt]{article}\usepackage[]{graphicx}\usepackage[]{xcolor}
\makeatletter
\def\maxwidth{ %
  \ifdim\Gin@nat@width>\linewidth
    \linewidth
  \else
    \Gin@nat@width
  \fi
}
\makeatother

\definecolor{fgcolor}{rgb}{0.345, 0.345, 0.345}
\newcommand{\hlsng}[1]{\textcolor[rgb]{0.192,0.494,0.8}{#1}}%
\newcommand{\hlopt}[1]{\textcolor[rgb]{0,0,0}{#1}}%
\newcommand{\hldef}[1]{\textcolor[rgb]{0.345,0.345,0.345}{#1}}%
\newcommand{\hlkwd}[1]{\textcolor[rgb]{0.737,0.353,0.396}{\textbf{#1}}}%

\usepackage{framed}
\makeatletter
\newenvironment{kframe}{%
 \def\at@end@of@kframe{}%
 \ifinner\ifhmode%
  \def\at@end@of@kframe{\end{minipage}}%
  \begin{minipage}{\columnwidth}%
 \fi\fi%
 \def\FrameCommand##1{\hskip\@totalleftmargin \hskip-\fboxsep
 \colorbox{shadecolor}{##1}\hskip-\fboxsep
     \hskip-\linewidth \hskip-\@totalleftmargin \hskip\columnwidth}%
 \MakeFramed {\advance\hsize-\width
   \@totalleftmargin\z@ \linewidth\hsize
   \@setminipage}}%
 {\par\unskip\endMakeFramed%
 \at@end@of@kframe}
\makeatother

\definecolor{shadecolor}{rgb}{.97, .97, .97}
\definecolor{messagecolor}{rgb}{0, 0, 0}
\definecolor{warningcolor}{rgb}{1, 0, 1}
\definecolor{errorcolor}{rgb}{1, 0, 0}
\newenvironment{knitrout}{}{} 

\usepackage{alltt}

\usepackage{natbib}
\usepackage{usebib} 
\bibinput{bibliography}

\usepackage{bm}
\usepackage{amsmath}
\usepackage{numprint}
\usepackage{mathtools}
\mathtoolsset{showonlyrefs} 

\usepackage{color}
\usepackage[color=green!20!white,textsize=scriptsize,textwidth=28mm]{todonotes}

\usepackage[utf8]{inputenc}
\usepackage{url}
\usepackage[printonlyused]{acronym}
\usepackage{enumerate}
\usepackage{setspace}
\usepackage[left=2cm,top=2cm,right=3.2cm,bottom=3cm,bindingoffset=0.5cm]{geometry}
\usepackage{titlesec}
\usepackage{lineno}
\usepackage{pbox}
\usepackage{multirow}
\usepackage{booktabs}

\usepackage{algorithm}
\usepackage{algorithmic}
\usepackage{caption}

\usepackage{titling}
\pretitle{\begin{flushleft}\LARGE\bfseries\sffamily}
\posttitle{\end{flushleft}}
\preauthor{\begin{flushleft}\large}
\postauthor{\end{flushleft}}
\predate{\begin{flushleft}\large}
\postdate{\end{flushleft}}
\titleformat*{\section}{\Large\bfseries\sffamily}
\titleformat*{\subsection}{\large\bfseries\sffamily}
\titleformat*{\subsubsection}{\bfseries\sffamily}
\titleformat*{\paragraph}{\bfseries\sffamily}
\usepackage[affil-it]{authblk} 

\renewcommand{\P}{\operatorname{\mathsf{Pr}}} 

\def\logit{\text{logit}}
\DeclareMathOperator{\E}{\mathsf{E}} 

\newcommand{\Bern}{\text{Bernoulli}}

\def\given{\,|\,}

\def\mm#1{\ensuremath{\boldsymbol{#1}}} 
\definecolor{grey}{rgb}{.6,.6,.6}
\newcommand{\N}{\mathcal{N}} %

\newcommand{\eg}{\textit{e.\,g.}}
\newcommand{\ie}{\textit{i.\,e.,}}
\acrodef{INLA}{Integrated nested Laplace approximations}
\acrodef{LGM}{latent Gaussian models}
\acrodef{GMRF}{Gaussian Markov random field}
\acrodef{MCMC}{Markov chain Monte Carlo}
\acrodef{GLMM}{Generalized linear mixed models}
\acrodef{GLM}{generalized linear model}
\acrodef{MC}{misclassification error}
\acrodef{PC}{penalized complexity}
\acrodef{ML}{Maximum Likelihood}

\title{Bayesian models for missing and misclassified variables using integrated nested Laplace approximations}

\author[1,*]{Emma Skarstein}
\author[2]{Leonardo Bastos}
\author[3]{H\aa vard Rue}
\author[1]{Stefanie Muff}

\affil[1]{Department of Mathematical Sciences, Norwegian University of Science and Technology, Norway}
\affil[2]{PROCC, Residência Oficial, Av. Brasil, 4365 - Manguinhos, Rio de Janeiro, RJ, Brazil}
\affil[3]{CEMSE Division, King Abdullah University of Science and Technology, Thuwal 23955-6900, Saudi Arabia}
\affil[*]{Corresponding author. E-mail: emma@skarstein.no}

\date{\small\today}
\IfFileExists{upquote.sty}{\usepackage{upquote}}{}
\begin{document}

\maketitle




\begin{abstract}
  Misclassified variables used in regression models, either as a covariate or as the response, may lead to biased estimators and incorrect inference. Even though Bayesian models to adjust for misclassification error exist, it has not been shown how these models can be implemented using integrated nested Laplace approximation (INLA), a popular framework for fitting Bayesian models due to its computational efficiency. Since INLA requires the latent field to be Gaussian, and the Bayesian models adjusting for covariate misclassification error necessarily introduce a latent categorical variable, it is not obvious how to fit these models in INLA. Here, we show how INLA can be combined with importance sampling to overcome this limitation. We also discuss how to account for a misclassified response variable using INLA directly without any additional sampling procedure. The proposed methods are illustrated through a number of simulations and applications to real-world data, and all examples are presented with detailed code in the supporting information.
\end{abstract}

\section{Introduction}




Misclassified variables arise when the recorded value of a categorical variable is different from the correct category. The error mechanism can be described by some misclassification probabilities, which are typically described by a misclassification matrix. The misclassified variable might be used either as a covariate or as a response variable in a model. For instance, we might be interested in the effect of exposure to some substance of toxin, say, smoking. This exposure may be observed with some misclassification. In the case of smoking it may be due to inaccurate reporting, but in other cases this could also be due to inaccurate tests, or other reasons. In this scenario, the misclassification would be in the variable used as a covariate. Similarly, we might wish to model the relationship between a person having a disease and certain risk factors, but this disease may only be diagnosed through an imprecise test. If we use the test result as the response data directly, without accounting for the misclassification, we are not capturing the actual relationship between the underlying disease status and the risk factors, and the estimated coefficients will not represent the true relationship we are interested in.

A large body of literature exists that studies the effect of misclassification error in different modelling scenarios. \citet[][Chapter 3]{gustafson2003} gives a comprehensive introduction to the effects of covariate misclassification in regression models. The book by \citet{yi2017} gives details for dealing with misclassified covariates as well as continuous measurement error in certain specific models, such as survival models, longitudinal models, and more. Another overview covering the effects of error in both continuous and categorical variables can be found in \citet{stratos1_2020}, an article that is written as a part of the STRATOS initiative, and nicely summarizes the current state of the field.
In a simple linear regression model, for example, one can easily show that nondifferential misclassification (that is, misclassification that does not depend on the response variable) in a binary covariate will lead to attenuation of the slope estimated using the misclassified variable, meaning that that coefficient will be underestimated compared to the slope using the correctly classified variable. This attenuation will be directly related to the misclassification probabilites (or equivalently, the sensitivity and specificity), as well as the probability of the correct variable being positive. In fact, a case with a sensitivity and specificity of 0.9, which could be interpreted as 10\% misclassification, gives 20\% attenuation, while a 10\% continuous measurement error only gives 1\% attenuation \citep{gustafson2003}. In cases where the true variable is much more likely to take one value than the other, the attenuation may be even more severe for the same sensitivity and specificity. It is clear that even with small misclassification probabilities, it is important to evaluate the effect of the misclassification on the produced results. Of course, many statistical models will be more complex than this simple linear regression scenario, and many articles detail the effects of covariate misclassification on different models \citep{greenland1980, readechristopher_kupper1991, buonaccorsi_etal2005}.

A number of publications have described various methods for adjusting model estimates for biases due to covariate misclassification. One popular method is the misclassification simulation extrapolation approach \citep[MC-SIMEX,][]{kuchenhoff_etal2006}, where more misclassification is added artificially in order to examine the behavior of the estimate under more extreme misclassification, and then an adjusted estimate is obtained by extrapolating backwards to the case with no misclassification. This approach is also used by \citet{rutkowski_zhou2015}. Moreover, a number of previous approaches have adapted a Bayesian method. Using a Bayesian approach, the data generating process can be mirrored through a hierarchical framework where priors, which may be informed by some additional information one may have about these, can be set to ensure the identifiability of the model. Most of this literature covers binary misclassification \citep{gustafson2003, ren_stone2007, chu_etal2010, xia_gustafson2016, hogg2018}, but \citet{nelson_etal2018} discuss a Bayesian model for misclassification in count variables. Additionally, some Bayesian models account for both misclassification and missingness in categorical variables \citep{luta_etal2013, xia_akakpo2022}, while others allow for both categorical and continuous measurement error along with missingness \citep{goldstein_etal2018}.

Similarly, response misclassification error leads to more severe biases than what is typical for continuous response variables. If a continuous response variable has homoscedastic additive measurement error, this error is absorbed by the residual term, and does not typically lead to any biases in the model parameters, though it will increase uncertainty \citep[see \eg][]{carroll_etal2006}. With categorical response misclassification the same is not true, however, and one must therefore be careful when dealing with a misclassified response variable. Some early work describes the effect of binary response misclassification \citep{copeland_etal1977, hausman_etal1998, neuhaus1999}. There is also some literature considering ordinal response misclassification in a Bayesian setting \citep{mwalili_etal2005, naranjo_etal2015}, while others describe binary response misclassification in a Bayesian setting \citet{paulino_etal2003, bollinger_vanhasselt2017}. Binary regression with misclassification in both the response and covariates has also been considered in a few publications \citep{tang_etal2015, liu_etal2016}. \citet{hofler2005} gives a review of the field before 2005, and covers both covariate and response misclassification. 

\ac{INLA} provide an efficient framework for fitting Bayesian models, conveniently available through the R-package R-INLA \citep{rue_etal2009}. The approach is significantly faster than sampling based methods like Markov chain Monte Carlo, which is especially convenient when working with more complex Bayesian models where the running time needed to fit the model can pose a practical limitation in the workflow. In the case of measurement error, it has previously been shown how to model continuous error in \ac{INLA} \citep{muff_etal2015}, and how to extend this to also account for missing data or variables with multiple types of continuous error \citep{skarstein_etal2023}. However, \ac{INLA} requires the latent effects to be a Gaussian Markov random field (GMRF), which restricts the applications slightly. When the latent mismeasured variable is continuous, it can be modeled using a Gaussian distribution \citep{muff_etal2015}. On the other hand, in the case when a covariate is categorical and is known to be observed with some misclassification probabilities, accounting for this misclassification necessitates the introduction of a discrete latent variable, which is not compatible with the latent Gaussian field.

To overcome this limitation, we propose to combine INLA with importance sampling, in order to account for misclassification error. This builds on previous work \citep{berild_etal2022}, which has shown how to combine \ac{INLA} with importance sampling, expanding the class of models that can be fit in \ac{INLA}. In our proposed approach, \ac{INLA} is used to obtain marginal likelihoods of the regression model conditioned on the latent categorical variable for each instance of the sampled latent variable. The respective marginal likelihood is used to weight the marginal posterior distribution of the regression parameters via Monte-Carlo integration.
As an alternative approach, we also show how an interpretation of the misclassified variable as a discretized error prone continuous variable can be used to adjust for misclassification using INLA alone. This leads to a much faster approach. The validity of this interpretation is dependent on the situation, and we discuss the details of this.


Technically, the problem of misclassified categorical variables is very similar to that of missing categorical variables, since both scenarios require specifying a model that introduces a latent variable describing the distribution of the correct version of the variable with misclassification. \citet{skarstein_etal2023} took advantage of this connection in the context of continuous variables with measurement error and/or missingness in INLA, and correspondingly we will also show how the misclassification model can be used to impute missing values of the categorical variable.

As mentioned, another type of misclassification concerns the case when a variable is used as the response in the model rather than when it is a covariate. As we will show, it is then equally important to account for the misclassification as in the covariate case. In contrast to the model with a misclassified covariate, we can write this model without the need for a latent discrete variable, meaning that it can technically be fit using only R-INLA.  We show an example of how adjustment for response misclassification can be done in the scenarios where it is suitable, and also suggest a way to proceed when the sensitivity and specificity are not known precisely. Unfortunately this becomes numerically unstable in R-INLA, especially for low values of sensitivity and specificity (corresponding to high misclassification probabilities), meaning that this approach must be used with caution.

In Section \ref{sec:mc}, we begin with a formal description of misclassification, and introduce the notation used in the rest of the paper. Next, Section \ref{sec:cov_mc} discusses how to treat covariate misclassification using INLA within importance sampling, as well as describing the special case where the misclassified variable is a discretized continuous variable, and explaining how to do missing covariate imputation for discrete covariates. Response misclassification in INLA is covered in Section \ref{sec:response_misclass}. We then demonstrate the validity of the proposed methods in a few different simulated scenarios in Section \ref{sec:simulations}. Section \ref{sec:applications} illustrates how the methods can be used on real-world data, and what considerations need to be taken in order to tailor the models to the data at hand. Finally, we discuss the usefulness of the methods and other aspects in the Section \ref{sec:discussion}.

\begin{table}[]
    \centering
    \caption{Description of notation used in this paper.}
    \begin{tabular}{cp{4.5in}}
    \toprule
        Notation & Description\\
        \midrule
        $n$ & Number of observations. \\
        $p$ & Number of covariates. \\
        $\bm{w} = (w_1, \dots, w_n)^\top$ & The misclassified, observed version of a categorical covariate. In this paper we mostly consider the binary case where $w_i \in\{0,1\}$. \\
        $\bm{x} = (x_1, \dots, x_n)^\top$ & The correct (latent) version of the categorical covariate with misclassification. \\
        $\bm{s} = (s_1, \dots, s_n)^\top$ & The misclassified, observed version of a categorical response. In this paper we mostly consider the binary case where $s_i \in\{0,1\}$.\\
        $\bm{y} = (y_1, \dots, y_n)^\top$ & The correct (but potentially latent) version of a response. \\
        $\bm{Z}$ & An $n\times (p-1)$ matrix where each column is any covariate in the main model of interest that are assumed to not have error (in addition to $\bm{x}$). \\
        $\widetilde{\bm{Z}}$ & An $n\times (p-1)$ matrix where each column is any covariate in the model for the variable with misclassification (the exposure or imputation model) that are assumed to not have error. \\
        $\mathsf{M}$ & The misclassification matrix. \\
        $\pi_{kl}$ & The entries of the misclassification matrix, $\pi_{kl} = \Pr(w_i = k \given x_i = l)$, in the binary case $\pi_{01}$ and $\pi_{10}$ are the misclassification probabilities, while $\pi_{00}$ and $\pi_{11}$ are the specificity and sensitivity, respectively. \\
        $\bm{\beta}$ & Any coefficients in the main model of interest. \\
        $\bm{\alpha}$ & Any coefficients in the exposure/imputation model. \\
        $\bm{p}_x = (p_{x_1}, \dots, p_{x_n)^\top}$ & Success probabilities for the latent covariate, $p_{x_i} = \Pr(x_i = 1)$. If there are no covariates in the exposure model, only an intercept, then $p_{x_i}$ will be the same for all $i$. Otherwise, this probability will be dependent on some observed covariates. \\
        $\bm{p}_y = (p_{y_1}, \dots, p_{y_n)^\top}$ & Success probabilities for a latent binary response, $p_{y_i} = \Pr(y_i = 1)$. If there are no covariates in the model, only an intercept, then $p_{y_i}$ will be the same for all $i$, but more commonly this will depend on some covariates.\\
        \bottomrule
    \end{tabular}
    \label{tab:notation}
\end{table}

\section{What is misclassification error?}\label{sec:mc}

For a binary variable $\bm{x} = (x_1, \dots, x_n)^\top$, assume we only observe a misclassified version $\bm{w} = (w_1, \dots, w_n)^\top$, where the error mechanism can be described by the misclassification matrix
\begin{equation} \label{eq:mcmatrix}
  \mathsf{M}=
  \left(
  \begin{matrix}
  \pi_{00} & \pi_{10} \\
  \pi_{01} & \pi_{11} \\
  \end{matrix}
  \right) \ ,
\end{equation}
with $\pi_{10} = \P(w_i=1 \ \given  x_i=0)$ and $\pi_{01} = \P(w_i=0 \given  x_i=1)$, for a given observation $i$. Then we can write the distribution of the observations as
\begin{align}
  w_i \given (x_i = l) &\sim \Bern(\Pr(w_i = 1 \given x_i = l)) \label{eq:mcdist} \\& \sim \Bern(\pi_{1l}) \ ,
\end{align}
where $l$ indicates the two possible values for $x_i$, $l \in \{0,1\}$. Since $\pi_{10} = 1-\pi_{00}$ and $\pi_{01} = 1-\pi_{11}$, the misclassification is often described using the sensitivity (true positive rate) $\pi_{11}$ and specificity (true negative rate) $\pi_{00}$, though specifying the two misclassification probabilities $\pi_{10}$ and $\pi_{01}$ is just as valid.
This concept can be easily generalized to a categorical variable with $k$ levels, in which case $\mathsf{M}$ is a $k\times k$ matrix, and the observed values come from a categorical (generalized Bernoulli) distribution.

One useful and straightforward generalization is to allow the misclassification probabilities to depend on an additional variable $\bm{z}$. In the binary case, for example, $\pi_{10}$ and $\pi_{01}$ could be modelled as
\begin{equation}
  \logit(\pi_{10}) = \gamma_{00} + \gamma_{0z}\bm{z}   \quad
  \text{and}  \quad
  \logit(\pi_{01}) = \gamma_{10} + \gamma_{1z}\bm{z} \ ,
\end{equation}
with regression parameters $\gamma_{00}$, $\gamma_{0z}$, $\gamma_{10}$, and $\gamma_{1z}$ \citep[][p.~70]{yi2017}, or equivalent models could be specified for $\pi_{00}$ and $\pi_{11}$ instead.

Often, it is assumed that covariate misclassification is \emph{nondifferential}, meaning that the misclassified covariate is conditionally independent of the response given the true covariates. This means that the misclassification probabilities do not change depending on the response value. In some cases this assumption may not hold however, in which case different misclassification probabilities can be assumed to depend on the value of the response. If the misclassification is differential, it is even more difficult to predict the effects of the misclassification, and it becomes very difficult to estimate the parameters unless the true value for the misclassified variable is observed for some entries \citep{carroll_etal2006}. We will consider an example of this later on in this paper.

In the remainder of this paper, we will use the notation $\bm{x}$ and $\bm{w}$ when referring to the correctly and misclassified versions of a covariate, respectively, while we will use $\bm{y}$ and $\bm{s}$ to refer to the correct and misclassified versions of a response variable. The general misclassification mechanism is still the same in both cases, the only difference is how they are then used in the larger modelling context. For an overview of this and other notation used in this paper, see Table \ref{tab:notation}.



\section{Accounting for covariate misclassification error}\label{sec:cov_mc}

\subsection{A model for covariate misclassification}\label{sec:cov_mc_model}

Assume we have $n$ observations in a generalized linear model (GLM).
The data are given as $(\bm{y},\bm{Z},\bm{x})$, with
$\bm{y}=(y_{1},\ldots,y_n)^\top$ denoting the response,
$\bm{Z}$ a covariate matrix of
dimension $n \times (p-1)$ for $p-1$ error-free covariates, and
$\bm{x}=({x}_1,\ldots,{x}_n)^\top$ an error-prone binary covariate
whose true values are unobservable, but instead we observe $\bm{w}=({w}_1,\ldots,{w}_n)^\top$. 
Suppose $\bm{y}$ is of exponential family form with mean $\mu_{i}= \E(y_{i}\given {x_i})$, linked to the linear predictor $\eta_{i}$ via
\begin{align}
  \mu_{i} &= h(\eta_{i}) \ , \nonumber \\
 \eta_{i} &= \beta_0 + \beta_x x_{i} + \bm{Z}_i^\top\bm\beta_z  + f(\cdot)\ ,\label{eq:glmb}
\end{align}
with $f(\cdot)$ denoting unknown functions of the covariates or additional unstructured terms.

Let the misclassification matrix $\mathsf{M}$ for $\bm{x}$ be given by Equation \eqref{eq:mcmatrix}, meaning that the \emph{error model} for $\bm{w}$ would be a Bernoulli distribution conditional on the correct value of the covariate (as described in Equation \eqref{eq:mcdist}),
\begin{align}
  w_i \given (x_i = l) &\sim \Bern(\pi_{1l}),
\end{align}
for $l \in \{0,1\}$,
and let us assume that the observed variable $\bm{x}$ depends on covariates $\widetilde{\bm{Z}}$ (which could be a subset of the covariates in $\bm{Z}$) according to some \emph{exposure model} $\pi(\bm{x}\given \bm{Z})$ \citep{richardson_gilks1993B, richardson_gilks1993A}. The latter specifies the distribution of the unknown variable (\eg, a risk factor) in the data. A standard choice would be
\begin{equation}\label{eq:exposure}
  \logit[\E( \bm{x} \given \widetilde{\bm{Z}})] =  \alpha_0 \mm{1} + \widetilde{\bm{Z}}\bm\alpha_z \ ,
\end{equation}
with intercept $\alpha_0$, $n\times 1$ vector $\mm{1}$ and slope vector $\bm{\alpha}_z$ \citep[see \eg,][Chapter 5]{gustafson2005}. This model could of course be adapted or extended to reflect the data at hand, for instance by adding interaction effects or random effects.


\subsection{Covariate misclassification error in INLA}

Since the correct version of a mismeasured variable $\bm{x}$ is unobserved, it is represented as a latent variable in the model. In INLA, all latent variables are assumed to be Gaussian. In the case where the mismeasured variable is continuous, this does not usually pose a problem, as it can then typically be assumed to be Gaussian \citep{muff_etal2015}. However, in the discrete case, when the misclassified variable follows a Bernoulli or categorical distribution, the correct version of the variable cannot be included in the latent field, which makes it unfeasible to directly fit these models with INLA. The described issue relating to the discrete latent variable can be circumvented by combining INLA with importance sampling \citep{berild_etal2022} or MCMC \citep{gomezrubio_rue2018}, in which the model conditioned on the misclassified variable is fit with INLA, while the latent version of the misclassified variable is estimated outside of INLA.

Combining INLA with importance sampling or MCMC both have advantages and disadvantages. In importance sampling, each iteration is independent of the next, meaning that iterations can be run in parallel. However, if we also wish to estimate the coefficients in the logistic regression model describing the latent variable, or the misclassification probabilities themselves, then an MCMC approach must be taken, since the model for estimating the latent covariate $\bm{x}$ will depend on the output from the previous iteration.
We also present an alternative approach, where the misclassified binary variable is assumed to be a discretization of an underlying Gaussian variable. In that case, the entire model can be fit using only INLA.

\subsubsection{Accounting for covariate misclassification with INLA and importance sampling}\label{sec:is_mc}
Assuming the scenario described in Section \ref{sec:cov_mc_model}, we adapt an approach using importance sampling to sample the latent binary covariate, while the remaining model conditioned on this latent variable is fit using R-INLA \citep{berild_etal2022}. In this section, we assume the parameters of the exposure model and the misclassification probabilities to be known.

Let the set of all regression parameters from the main model of interest be denoted by $\bm{\theta}_y$.
To obtain marginal posterior distributions of the regression parameters $\bm{\theta}_y$, we use that
\begin{equation}\label{eq:mar_post}
  \pi(\bm{\theta}_y \given \bm{y}, \bm{w}, \bm{Z}) = \int \pi(\bm{\theta}_y \given \bm{x},  \bm{y}, \bm{Z})    \pi(\bm{x} \given \bm{y}, \bm{w}, \bm{Z}) d\bm{x} \ ,
\end{equation}
and that 
\begin{align*}
  \pi(\bm{x} \given \bm{y}, \bm{w}, \bm{Z}) \propto \pi(\bm{x} \given \bm{w}, \bm{Z}) \pi(\bm{y} \given \bm{x}, \bm{Z}) \ .
\end{align*}
%
If the model $\pi(\bm{x}\given\bm{Z})$ for the erroneous covariate $\bm{x}$ and the error model $\pi(\bm{w}\given \bm{x})$ are known, it is possible to sample from $ \pi(\bm{x} \given \bm{w}, \bm{Z}) \propto \pi(\bm{w}\given \bm{x}) \pi(\bm{x}\given \bm{Z})$. If $\pi(\bm{w}\given\bm{x}) = \Bern(\bm{p}_{w\given x})$ where $\bm{p}_{w\given x}$ is decided by the value of $\bm{x}$ according to the misclassification matrix, and $\pi(\bm{x}\given\bm{Z}) = \Bern(\bm{p}_x)$ where $\logit(\bm{p}_x) = \alpha_0\bm{1} + \bm{Z}\bm{\alpha}_z$, then $\pi(\bm{x}\given \bm{w}, \bm{Z})$ is also a Bernoulli distribution with success probability
\begin{align}\label{eq:sampling_prob}
  p_{x_i\given w_i = k, Z_i} &= \Pr(x_i = 1 \given w_i = k) \\[8pt]
  &= \frac{\Pr(x_i = 1, w_i = k)}{\Pr(w_i = k)} \\[8pt]
  &= \frac{\Pr(w_i = k \given x_i = 1)\Pr(x_i = 1)}{\sum_{l \in\{0,1\}}\Pr(w_i = k\given x_i = l)\Pr(x_i = l)} \\[8pt]
  &= \frac{\pi_{k1}p_{x_i}}{\pi_{k0}(1-p_{x_i}) + \pi_{k1}p_{x_i}} \ ,
\end{align}
which can be calculated for each observation $i$ since $\bm{w}$ and $\bm{Z}$ are observed. Since $\bm{p}_x$ is allowed to depend on $\bm{Z}$ it may vary for different individuals, and therefore $\bm{p}_{x\given w, Z}$ will vary both depending on $\bm{w}$ and $\bm{Z}$. For given samples $\bm{x}^{(j)}$ ($1\leq j \leq M$) sampled from  $ \pi(\bm{x} \given \bm{w}, \bm{Z})$ the marginal posterior distribution \eqref{eq:mar_post} can then be approximated by Monte-Carlo integration
\begin{equation}\label{eq:mar_post_app}
  \pi(\bm{\theta}_y \given \bm{y}, \bm{w}, \bm{Z}) \approx \sum_{j=1}^M \pi(\bm{\theta}_y \given \bm{x}^{(j)},  \bm{y}, \bm{Z}) \cdot \omega_j
\end{equation}
with weights proportional to the (conditional) marginal likelihood $\omega_j \propto \pi(\bm{y} \given \bm{x}^{(j)}, \bm{Z})$, normalized such that $\sum_j \omega_j=1$, where the marginal likelihood $\pi(\bm{y} \given \bm{x}^{(j)}, \bm{Z})$ is obtained from fitting the model with $\bm{x}^{(j)}$ as the covariate in \ac{INLA}.
As an example, the posterior mean of $\beta_x$ can be approximated as
\begin{equation*}
  \E(\beta_x  \given \bm{y}, \bm{w}, \bm{Z}) \approx \sum_{k=1}^K  \E(\beta_x \given \bm{x}^{(k)},\bm{y}, \bm{Z}) \cdot \omega_k  \ ,
\end{equation*}
and posterior $\alpha$-quantiles $q_\alpha(\beta_x)$ are given as
\begin{equation*}
  q_\alpha(\beta_x \given \bm{y}, \bm{w}, \bm{Z}) \approx \sum_{k=1}^K q_\alpha(\beta_x \given \bm{x}^{(k)},\bm{y}, \bm{Z})  \cdot  \omega_k \ .
\end{equation*}

\begin{algorithm}
    \caption{INLA within importance sampling for covariate misclassification adjustment}
    \begin{algorithmic}
    {\footnotesize
    \STATE Given observed data $\bm{y}$, covariate with misclassification $\bm{w}$ and correct version of this covariate $\bm{x}$, and any $p$ additional covariates $\bm{Z}$ measured without error.
    \FOR{$j = 1, 2, \dots, M$}
      \STATE Sample $\bm{x}^{(j)}$ from $\Bern(\bm{p}_{x\given w, Z})$, where $p_{x_i\given w_i = k, Z_i} = \frac{\pi_{k1}p_{x_i}}{\pi_{k0}(1-p_x) + \pi_{k1}p_{x_i}}$
      \STATE Fit the regression model of interest conditional on $\bm{x}^{(j)}$ with R-INLA, to approximate the conditional marginals $\pi(\bm{\theta}_y \given \bm{y}, \bm{x}^{(j)})$.
      \STATE From the R-INLA output, obtain the conditional means of $\bm{\theta}_y$, giving $\bm{\theta}_y^{(j)}$, as well as the approximated marginal likelihood $\tilde{\pi}(\bm{y}\given \bm{x}^{(j)}, \bm{Z})$.
      \STATE Save $\bm{x}^{(j)}$, $\bm{\theta}_y^{(j)}$ and the approximate marginal likelihood $\tilde{\pi}(\bm{y}\given \bm{x}^{(j)}, \bm{Z})$.
    \ENDFOR
    \STATE Calculate weights proportional to the marginal likelihood $\omega_j \propto \pi(\bm{y}\given \bm{x}^{(j)}, \bm{Z})$, $j = 1,\dots, M$, normalized such that $\sum_j \omega_j = 1$.
    \STATE The marginal posterior distribution can now be approximated by \\
    $$
    \pi(\bm{\theta}_y \given \bm{y}, \bm{w}, \bm{z}) \approx \sum_{j=1}^M \pi(\bm{\theta}_y \given \bm{x}^{(j)},  \bm{y}, \bm{z}) \cdot \omega_j
    $$
    \\
    }
    \end{algorithmic}
    \label{inla-is}
\end{algorithm}

The entire procedure is summarized in Algorithm \ref{inla-is}. In this approach, the coefficients $\bm{\alpha}$ of the model for the latent variable $\bm{x}$ are assumed to be known. In certain scenarios, such as when we have sufficient validation data or when the model only contains an intercept, this is a reasonable assumption. We could also have an intercept only model, if $\bm{x}$ does not depend on other covariates. Code implementing this method is available in the supporting information. The method has been implemented in an R-package, \texttt{inlamisclass}, which can be downloaded from GitHub at \url{https://github.com/emmaSkarstein/inlamisclass}.

\subsubsection{Covariate misclassification from a latent Gaussian variable with measurement error}\label{sec:mc_from_cont}
In many cases, a recorded binary variable can be interpreted as a discretization of a continuous variable, which may often be assumed to follow a normal distribution. Examples of this are medical tests, where the binary decision is based on a concentration of markers in the blood or other samples, or genetic threshold models (sometimes called liability threshold models) used in medicine and genetics \citep{felsenstein2005}. In such models, discrete traits, such as disease status, actually arise from an underlying continuous trait.

In the case where this countinuous trait is subject to measurement error before it is discretized, the resulting discrete variable will have misclassification which arises from a continuous classical measurement error mechanism \citep[see \eg,][for classical measurement error]{carroll_etal2006}. The resulting observed value will still be misclassified, but since the latent variable is Gaussian, this is a scenario in which it will be possible to fit a model purely using INLA.

To describe such a scenario precisely, we introduce new notation. Let $\bm{y}$ denote the response variable, as before, with mean $\bm{\mu} = \E(\bm{y} \given \bm{x}_c, \bm{Z})$, where we $\bm{x}_c$ is the underlying continuous latent variable, and $\bm{w}_c$ is the continuous variable plus some measurement error. Lastly, let $\bm{w}_d$ be the observed discretized version of $\bm{w}_c$. For the discussion, we also introduce the discretized version of $\bm{x}_c$, denoted $\bm{x}_d$, which would be the correctly classified binary variable. Then the setup can be summarized as follows:
\begin{align*}
\bm{\mu} &= h(\beta_0\bm{1} + \beta_{x_c}\bm{x}_{c} + \bm{Z}\bm{\beta} + f(\cdot)) \ , \\
\bm{w}_{c} &= \bm{x}_{c} + \bm{u} \ ,  & \bm{u} \sim \N(\bm{0}, \sigma_{u}^2\bm{I}) \ , \\
\bm{w}_{d} &= \text{I}(\bm{w}_{c} > 0) \ , & \\
\bm{x}_{c} &= \alpha_0\bm{1} + \widetilde{\bm{Z}}\bm{\alpha} + \bm{\varepsilon}_{x} \ ,  &\bm{\varepsilon}_{x} \sim \N(\bm{0}, \sigma_{\varepsilon_x}^2\bm{I}) \ , &
\end{align*}
where $\bm{Z}$ and $\widetilde{\bm{Z}}$ are both covariate matrices with whichever covariates are relevant to include, these are assumed to be without error. In the indicator function the cutoff can be set to zero without loss of generality, since if any other cutoff $c$ is used the indicator function can always be rephrased as $\text{I}(w_{c,i}-c > 0)$. Now, we have that
\begin{align*}
\Pr(w_{d,i} = 1 \given x_{c,i}) &= \Pr(w_{c,i} > 0) \\
&= \Pr(x_{c,i} + u_{i} > 0) \\
&= \Pr(u_i > -x_{c,i}) \\
&= \Pr\left(\frac{u_{i}}{\sigma_{u}} < \frac{x_{c,i}}{\sigma_{u}}\right) \\
&= \Phi\left(\frac{x_{c,i}}{\sigma_{u}}\right) \ ,
\end{align*}
since $u_{i}/\sigma_{u} \sim \mathcal{N}(0, 1)$, and $\Pr(u_i > -x_{c,i}) = \Pr(u_i < x_{c,i})$ since the distribution of $\bm{u}$ is centered at zero. This relation means that we can model $\bm{w}_d$ as a Bernoulli variable with a probit link function. We can re-write the model:
\begin{align*}
\bm{\mu} &= h(\beta_0\bm{1} + \beta_{x_c}\bm{x}_{c} + \bm{Z}\bm{\beta} + f(\cdot)) \ , \\
\bm{w}_{d} &\sim \Bern(\Phi(\bm{x}_c/\sigma_{u})) \ , & \\
\bm{x}_{c} &= \alpha_0\bm{1} + \widetilde{\bm{Z}}\bm{\alpha} + \bm{\varepsilon}_{x} \ ,  &\bm{\varepsilon}_{x} \sim \N(\bm{0}, \sigma_{\varepsilon_x}^2\bm{I}) \ , &
\end{align*}
The implementation of this in R-INLA closely follows the approach of \citet{muff_etal2015}, since this is essentially a classical measurement error model with an added layer of discretization.
An example using simulated data is shown in Section \ref{sec:cont_to_binary}, and code that illustrates how to implement this model in R-INLA is included in the supporting information.

\subsection{Missing values in categorical covariates in INLA}\label{sec:missing}
The Bayesian models for adjusting for covariate misclassification in Section \ref{sec:is_mc} and \ref{sec:mc_from_cont} can easily be adapted to impute missing values as well. In the latter case, since the latent variable is continuous, the method described in \citet{skarstein_etal2023} for accounting for missing observations in a continuous covariate in INLA can be used.

When using the INLA within importance sampling-approach outlined in Section \ref{sec:is_mc}, the proposed method can also easily be adapted to account for missing values. Conceptually, one could argue that missingness in a binary variable is equivalent to misclassification with a 50\% misclassification probability. \citet{skarstein_etal2023} utilize an analogous interpretation in the continuous case to impute continuous missing values in INLA. In both scenarios, the logic is that, as the misclassification probabilites (or the continuous measurement error variance) increase, we know less about the correct value of the observation. In the most extreme case, we have no information about the correct value of the covariate. In the continuous case, that would mean the measurement error variance approaching infinity, but of course it does not make sense for the misclassification probabilities to increase infinitely. Instead, the most extreme misclassification probability would be a misclassification probability of 50\%, meaning that the observed value gives us no information about the correct value of the covariate, and it might as well be missing.

If a given observation $i$ is missing for the covariate $\bm{w}$, then within the importance sampling setup we have that $\pi(x_i\given w_i, Z_i) = \pi(x_i \given Z_i)$, since $w_i$ is unobserved. Alternatively, using the interpretation of missingness as equivalent to 50\% misclassification, we reach the same conclusion. In the success probability in Equation \eqref{eq:sampling_prob}, we would have $\pi_{k1} = \pi_{k0} = 0.5$. This would give
\begin{equation}\label{eq:sampling_prob_missing}
  p_{x_i\given w_i = k, Z_i}
  = \frac{\pi_{k1}p_{x_i}}{\pi_{k0}(1-p_{x_i}) + \pi_{k1}p_{x_i}}
  = \frac{0.5p_{x_i}}{0.5(1-p_{x_i}) + 0.5p_{x_i}} = p_{x_i} \ ,
\end{equation}
leaving just $p_{x_i \given w_i, Z_i} = p_{x_i}$. That means that when the observations are missing, then the corresponding value for $x_i$ is just sampled directly from the exposure/imputation model. In the case when the covariate has missing values but no misclassification, then one only samples the values which are missing. A simulated example is provided in Section \ref{sec:missing_ex}.

\section{Accounting for response misclassification error}\label{sec:response_misclass}

In scenarios where we have a discrete response variable, misclassification in this response variable also leads to potentially severe biases in the model, and must be accounted for \citep[][Chapter 15.3]{magder_etal1997, carroll_etal2006}. In this section, we will only consider binary misclassification, defined by the misclassification matrix in the same way as for the covariate case, and the general setting described in Section \ref{sec:mc}. A typical scenario where this problem can arise is in epidemiology, when dealing with results from a test for a certain disease. The test for the disease will not be completely accurate, and there will be some misclassification of the disease status. In epidemiology, the misclassification is most often parameterised by the sensitivity and specificity. 

\subsection{A model for response misclassification} \label{sec:response_mc}
As mentioned before, in the misclassification terminology, the specificity will correspond to the entry $\pi_{00}$ in the misclassification matrix, whereas the sensitivity corresponds to the entry $\pi_{11}$. Let $\bm{y} = (y_1, \dots, y_n)^\top$ be the correct, but unobserved response (\ie \, the disease status), while $\bm{s} = (s_1, \dots, s_n)^\top$ is the observed version of $\bm{y}$ (\ie \, the test result). Then the model can be written in a similar way as for the case with covariate misclassification, that is,
\begin{align}
y_i &\sim \Bern(p_y) \ , \\
\logit(p_{y_i}) & = \beta_0 + \bm{Z}^\top_i\bm{\beta}_z \ ,\\
s_i\given(y_i = l) &\sim \Bern(\pi_{1l}) \ ,
\end{align}
where $\pi_{1l} = \Pr(s_i = 1 \given y_i = l)$, and $\bm{Z}_i$ is the $i$-th row in the matrix with all covariates. Then, the marginal distribution of $s_i$ is a Bernoulli distribution with success probability
\begin{align}\label{eq:p_s}
p_{s_i} &= \Pr(s_i = 1) \\
&= \Pr(s_i = 1\given y_i = 1)\cdot\Pr(y_i = 1) + \Pr(s_i = 1\given y_i = 0)\cdot\Pr(y_i = 0) \\
&= \pi_{11}p_{y_i} + (1-\pi_{00})(1-p_{y_i}) \ .
\end{align}
We wish to write the model in terms of $\bm{s}$, its respective success probability $p_{s_i}$ and the predictor $\beta_0 + \bm{Z}_i^\top\bm{\beta}_z$, while accounting for the sensitivity $\pi_{11}$ and specificity $\pi_{00}$. From the relation $p_{s_i} = \pi_{11}p_{y_i} + (1-\pi_{00})(1-p_{y_i})$ we can rearrange the expression, and get that
\begin{equation}
p_{y_i} = \frac{p_{s_i} -(1-\pi_{00})}{\pi_{11}-(1-\pi_{00})} \ .
\end{equation}
Using this new expression for $p_y$, the model for the observed variable $\bm{s}$ can instead be summarised as
\begin{align}
s_i &\sim \Bern(p_{s_i}) \ , \nonumber \\
\logit\left(\frac{p_{s_i} -(1-\pi_{00})}{\pi_{11}-(1-\pi_{00})}\right) &= \beta_0 + \bm{Z}^\top_i\bm{\beta}_z \ ,\label{eq:sslogit}
\end{align}
emphasizing that the model accounting for the sensitivity and specificity is in fact just a Bernoulli model with a different link function. The posterior distribution for $p_{y_i}$ is then
\begin{equation}
\pi(p_{y_i} \given \bm{s}, \pi_{00}, \pi_{11}) \propto p_{s_i}^{\sum_i y_i}(1-p_{s_i})^{n-\sum_i y_i}\pi(p_{y_i}) \ .
\end{equation}
In some scenarios, we may have good information on the values of the sensitivity and specificity, so that we can assume them to be known. However, if we do not know them exactly, but instead have an approximate idea of the values, for instance with 95\% probability intervals, we can integrate them out of the posterior distribution of $\bm{p}_{y}$, \ie
\begin{equation}
\pi(\bm{p}_{y} \given \bm{s}) = \int\int \pi(\bm{p}_{y}\given\bm{s}, \pi_{00}, \pi_{11})\pi(\pi_{00}, \pi_{11})d\pi_{00}d\pi_{11} \ .
\end{equation}
A numerical approximation for $\pi(\bm{p}_{y}\given\bm{s})$ can be obtained by summing through a grid of points chosen in the domain of $(\pi_{00}, \pi_{11})$ weighted by a joint probability distribution for them, that is,
\begin{equation}
\tilde\pi(\bm{p}_{y} \given \bm{s}) = \sum_k \pi\left(\bm{p}_{y}\given\bm{s}, \pi_{00}(k), \pi_{11}(k)\right)\omega(\pi_{00}(k), \pi_{11}(k)) \ ,
\end{equation}
where $\omega(\cdot,\cdot)$ is a weighting function such that $\sum_k \omega(\pi_{00}(k), \pi_{11}(k)) = 1$.

\subsection{Response misclassification in INLA}

In INLA, response misclassification does not pose the same technical difficulties as covariate misclassification, since the latent variable does not need to be explicitly modelled. The adjusted link function described in Equation \eqref{eq:sslogit} has been implemented in R-INLA as the link function \texttt{sslogit}. Unfortunately, when the sensitivity and/or specificity are large, the model suffers from numerical instabilities that can be difficult to diagnose. Because of this problem, the link function has not been made available by default. However, the supplementary material shows how this link function can be enabled. In Section \ref{sec:response_mc_example} we illustrate its use on a simulated example, where it is also shown how uncertainty about the sensitivity and specificity can be propagated into the model.

\subsection{Missing values in categorical responses in INLA}

For missing values in the response, the respective missing values can be predicted from their predictive distribution, and this is no different if the response is discrete or continuous. We therefore refer the reader to \citet[][Chapter 12.3]{gomezrubio2020} for a detailed description along with code showing how this is done in INLA.

\section{Simulations} \label{sec:simulations}

\subsection{Misclassification error in a covariate for linear regression} \label{sec:simstudy}
In this section, we look at a simple example with simulated datasets. To get an idea of the sensitivity of the proposed method, we simulated a new data set using the same parameters 10 times, and fit the same model to each data set, before examining the results. The data consists of three variables: the response $\bm{y}$, a misclassified binary covariate $\bm{w}$, and an error free continuous covariate $\bm{z}$. Additionally, we denote the correct, unobserved version of $\bm{w}$ as $\bm{x}$. First, each component of $\bm{z}$ was generated uniformly $z_i \sim \text{Unif}(-1,1)$ for $1\leq i \leq 100$, and then $\bm{x}$ was sampled according to an exposure model given as
\begin{equation}
  \text{logit}[E( \bm{x} \mid \bm{z})] =  \alpha_0 \mathsf{1} + \alpha_z\bm{z} \ ,
\end{equation}
with $\bm\alpha^\top=(\alpha_0, \alpha_z)=(-0.5,0.25)$.
This dependency was then also appropriately reflected in the analysis, assuming that $\bm{\alpha}$ was known. The misclassification is generated according to the misclassification matrix
\begin{equation}
  \mathsf{M}=
  \left(
  \begin{matrix}
  0.9 & 0.1 \\
  0.2 & 0.8 \\
  \end{matrix}
  \right) \ .
\end{equation}
The response $\bm{y}$ was simulated according to the linear model
\begin{equation}
  \bm{y} = \bm{1} + \bm{x} + \bm{z} + \bm{\epsilon} \ , \quad \bm{\epsilon} \sim\mathcal{N}(\bm{0}, \mathbf{I}) \ .
\end{equation}

\begin{figure}
  \includegraphics[width=\linewidth]{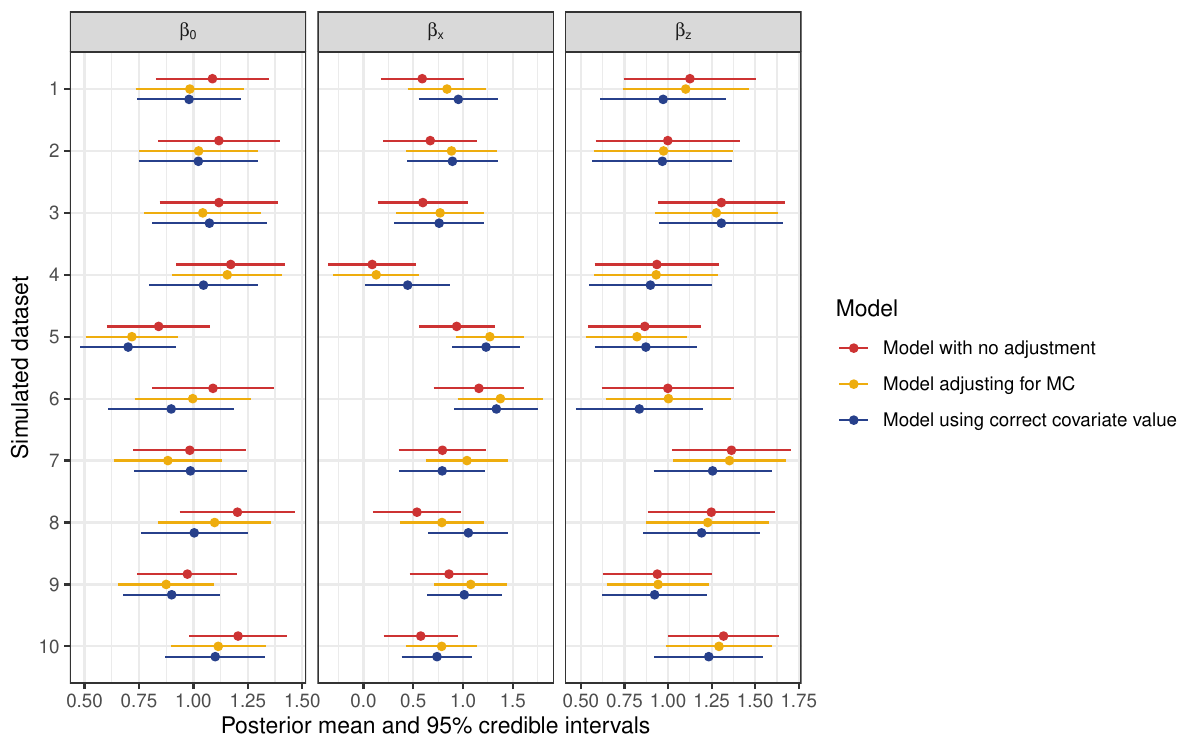}
  \caption{Posterior means and 95\% credible intervals for the estimated coefficients from ten different simulated examples, for the case where a covariate is subject to misclassification (MC).}
  \label{fig:simulation_study}
\end{figure}

\begin{figure}
  \includegraphics[width=\linewidth]{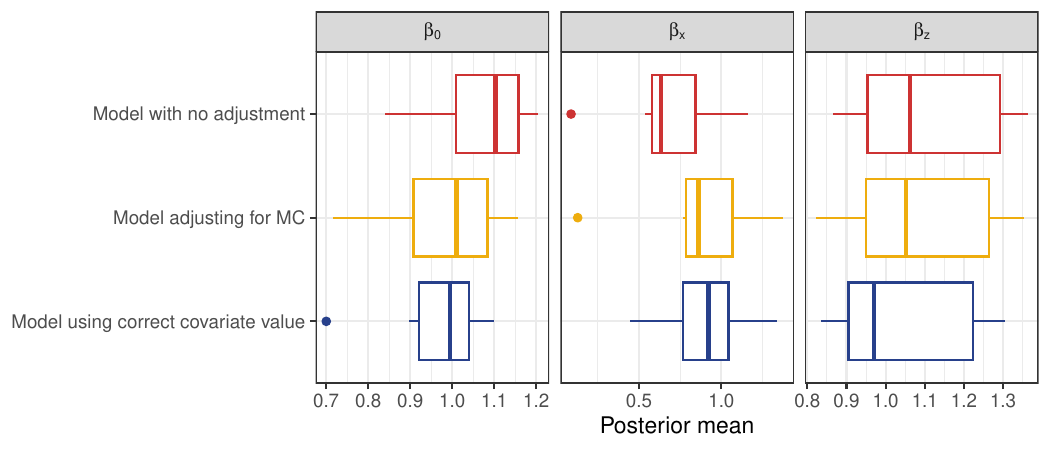}
  \caption{Box plots of the posterior means for the estimated coefficients from ten different simulated examples, for the case where a covariate is subject to misclassification.}
  \label{fig:simulation_boxplot}
\end{figure}

To account for the misclassification in the covariate, we fit a Bayesian model reflecting the data-generating process, using the proposed combination of INLA and importance sampling from Section \ref{sec:is_mc}. The procedure is run for \numprint{200000} iterations. The posterior means and 95\% credible intervals for $\beta_0$, $\beta_z$ and $\beta_x$ from each simulated data set can be seen in Figure \ref{fig:simulation_study}, where the results from fitting the model without adjusting for misclassification and the results from using the correct value for the covariate can also be seen. A summary across all ten data sets is given in Figure \ref{fig:simulation_boxplot}. With \numprint{200000} iterations, each model took us around seven hours to run.

\subsection{Misclassification from dichotomizing a mismeasured continuous covariate}\label{sec:cont_to_binary}

In this example, we simulate $n = 200$ observations from the general framework outlined in Section \ref{sec:mc_from_cont}. In particular, the variables are simulated as follows:
\begin{align}
  \bm{x}_c &\sim \N(\bm{0}, \bm{I}) \ , \\
  \bm{x}_d &= \text{I}(\bm{x}_c > 0) \ , \\
  \bm{w}_c &= \bm{x}_c + \bm{u} \ , \quad \bm{u} \sim \N(\bm{0}, \bm{I}) \ , \\
  \bm{w}_d &= \text{I}(\bm{w}_c > 0) \ ,
\end{align}
where $\bm{x}_c$ is the latent continuous version of the covariate, $\bm{x}_d$ is the latent discretized version of $\bm{x}_c$, $\bm{w}_c$ is the latent covariate with additive measurement error, and $\bm{w}_d$ is the discretized version of $\bm{w}_c$. Comparing the two discrete variables, $\bm{x}_d$ and $\bm{w}_d$, we can calculate the sample misclassification matrix, which in this case is
\begin{equation} \label{eq:mcmatrix_mc_from_cont}
  \mathsf{M}=
  \left(
  \begin{matrix}
  0.66 & 0.33 \\
  0.29 & 0.71 \\
  \end{matrix}
  \right) \ ,
\end{equation}
meaning that a significant proportion of the covariate has been misclassified. In this scenario, we assume that we only have access to $\bm{w}_d$ and the response $\bm{y}$, which we simulate as
\begin{equation}
  \bm{y} = \beta_0\bm{1} + \bm{x}_c + \bm{\varepsilon} \ ,
\end{equation}
with $\bm{\varepsilon} \sim \N(\bm{0}, \bm{I})$, $\beta_0 = 1$ and $\beta_{x_c} = 1$ .
To adjust for the misclassification that arises due to discretizing the mis-measured variable $\bm{w_c}$, we include an error model which describes how $\bm{w}_d$ follows a Bernoulli distribution with a probit link function:
\begin{align*}
\bm{y} &= \beta_0\bm{1} + \beta_{x_c}\bm{x}_{c} + \bm{\varepsilon} \ , \\
\bm{w}_{d} &\sim \Bern(\Phi(\bm{x_c}/\sigma_{u})) \ , & \\
\bm{x}_{c} &= \alpha_0\bm{1} + \widetilde{\bm{Z}}\bm{\alpha} + \bm{\varepsilon}_{x} \ ,  &\bm{\varepsilon}_{x} \sim \N(\bm{0}, \sigma_{\varepsilon_x}^2\bm{I}) \ , &
\end{align*}
where $\sigma_u$ is fixed to 1, $\beta_{x_c}$ is given a $\N(0, 1000)$ prior, and the remaining parameters are given default priors, see the supporting information. Since $\bm{w}_d$ is observed and the latent variable is $\bm{x}_c$, which is Gaussian, this model can be fit entirely in R-INLA, as described in Section \ref{sec:mc_from_cont}. The code is available in the supporting information. The model is set up similar to the continuous measurement error model described in \citet{muff_etal2015}, since we are essentially adjusting for a classical measurement error mechanism with the added layer of the discretization. The joint model is most easily specified using \texttt{inla.stack}, which enables us to specify each sub-model separately. The \texttt{copy} feature is used when specifying the formula to ensure that $\bm{x}_c$ is included correctly through all three levels of the model.

\begin{figure}
  \includegraphics[width=\linewidth]{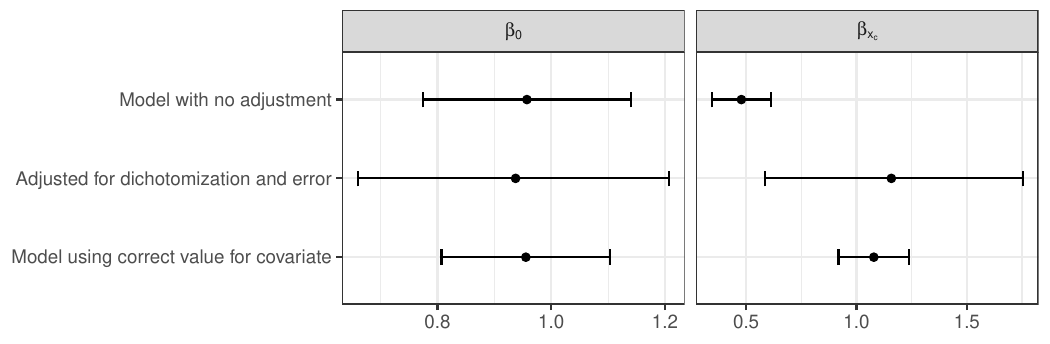}
  \caption{Posterior means and 95\% credible intervals for the example where the misclassification arises due to dichotomization of a continuous variable with error.}
  \label{fig:cont_to_binary}
\end{figure}

In order to evaluate the fit of this model, we compare the estimated coefficients to those from the model using the correct version of $\bm{x}_c$, and the model using the covariate with error, $\bm{w}_c$. Since $\bm{w}_d$ is the observed variable, one might want to compare the estimated coefficient to that estimated when $\bm{w}_d$ is used directly, but since the interpretation changes when the covariate is discrete rather than continuous, we do not make this comparison. We find that the model using the error prone continuous covariate dramatically underestimates the coefficient (which is expected for this kind of continuous measurement error), while the model that adjusts for the error better matches the correct coefficient (Figure \ref{fig:cont_to_binary}). The model that adjusts for the error also has quite large uncertainty, which is appropriate here, since the discretization process means that we loose a lot of the information present in the latent continuous variable.

\subsection{Missing observations in a binary covariate}\label{sec:missing_ex}
\begin{figure}
  \includegraphics[width=\linewidth]{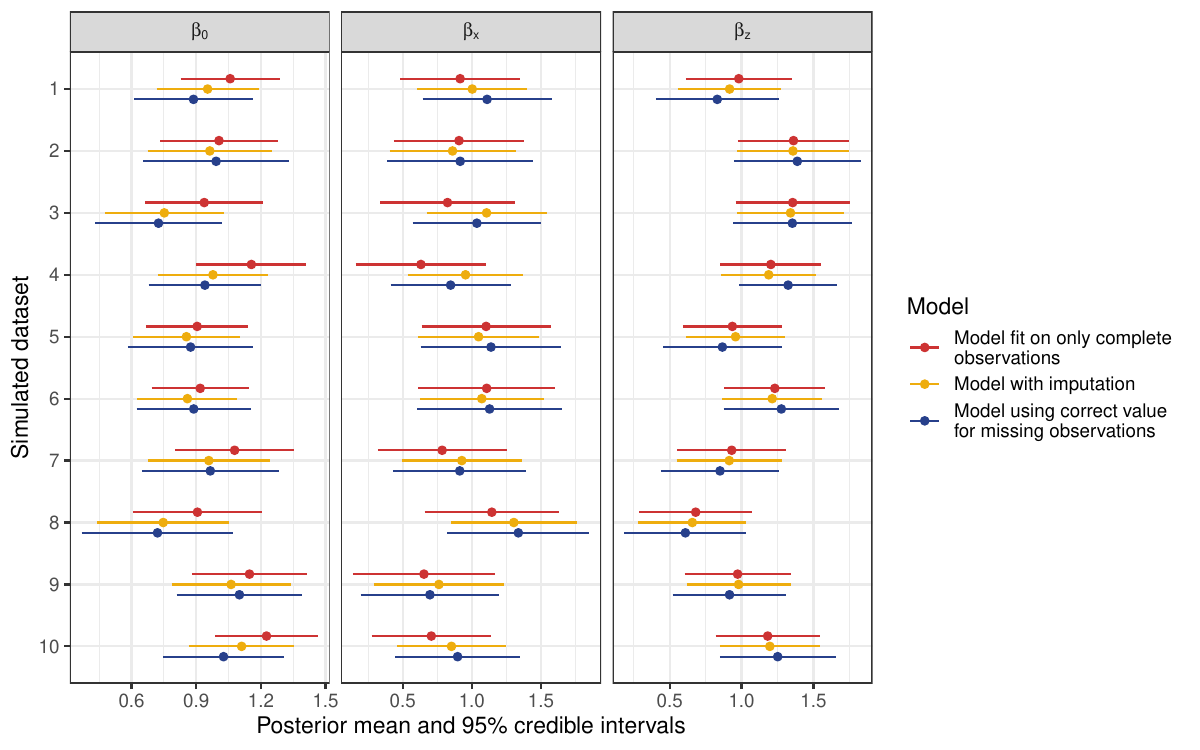}
  \caption{Posterior means and 95\% credible intervals for the estimated coefficients from ten different simulated examples, for the case where a covariate has missing observations.}
  \label{fig:missing_simulated}
\end{figure}

To illustrate the approach to handle missing data in a binary covariate, as proposed in Section \ref{sec:missing}, we simulate a binary covariate with 20\% missing values, but no misclassification. We use the same data simulation setup as the linear regression with a misclassified covariate simulated in Section \ref{sec:simstudy}, but instead of $\bm{w}$ being the misclassified version of $\bm{x}$, it has 20\% of the observations missing completely at random and no misclassification in the remaining observations. We fit the same main model as in Section \ref{sec:simstudy}, but without the misclassification model, as described in Section \ref{sec:missing}. We ran the importance sampling procedure for \numprint{100000} iterations. Details around how to do this technically can be found in the supporting information. The resulting posterior means and 95\% credible intervals for the coefficients of interest can be seen in Figure \ref{fig:missing_simulated}. Here, the results are also compared to the model that uses only complete cases, meaning that the observations with missingness are left out completely, and to a model fit using the correct version of the covariate without missing observations.

\subsection{Response misclassification}\label{sec:response_mc_example}
In this example, we simulate a misclassified response $\bm{s}$ to illustrate the effects of response misclassification and how it can be adjusted for using R-INLA. Let $\bm{y}$ be the latent response, for instance the disease status, while $\bm{s}$ is the misclassified response, which might be the test result indicating whether the patient has the disease. In this imagined scenario, the probability of having the disease is $p_y = 0.10$, and the test has sensitivity $\pi_{11} = 0.95$ and specificity $\pi_{00} = 0.90$.
Following the expression in Equation \eqref{eq:p_s}, we can simulate $n = 1000$ observations of $\bm{s}$ directly from
\begin{equation}
  \bm{s} \sim \Bern(p_s) \ ,
\end{equation}
where $p_s$ depends on $p_y$ and the sensitivity and specificity through
\begin{equation}
  p_s = \pi_{11}p_y + (1-\pi_{00})(1-p_y) \ .
\end{equation}

\begin{figure}
  \includegraphics[width=\linewidth]{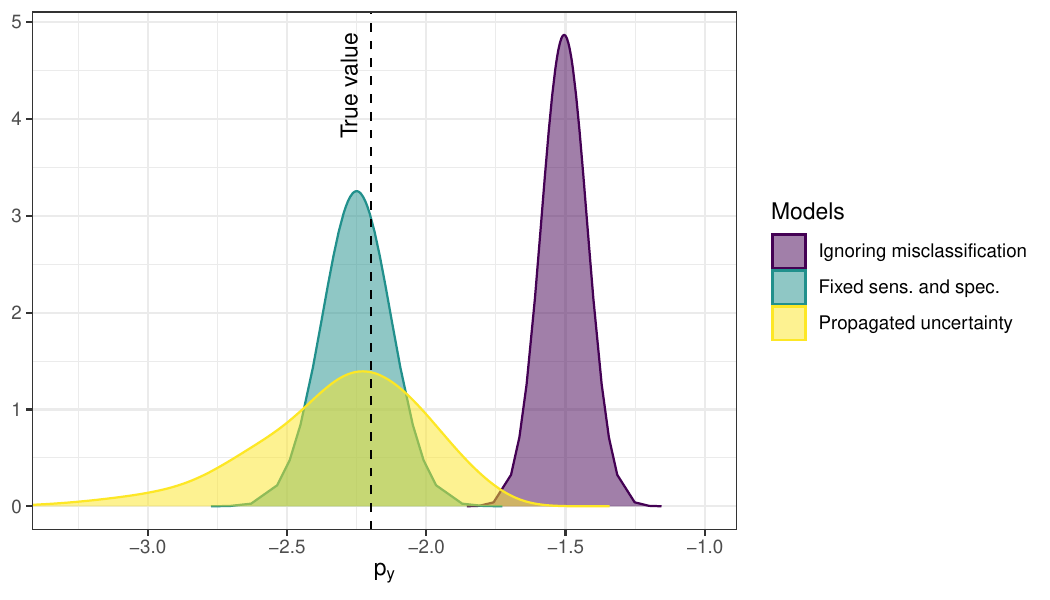}
  \caption{Posterior marginal distributions of the response success probability for models adjusting for misclassification (yellow and green) versus one that makes no adjustment (purple). The green model uses fixed values for sensitivity and specificity, whereas the yellow model incorporates the uncertainty in the sensitivity and specificity.}
  \label{fig:response_mc}
\end{figure}

We then fit three different models: first, we fit the model that does not account for the misclassification by simply using the regular link function.  Next we fit a model which assumes fixed sensitivity (95\%) and specificity
(90\%) and accounts for the misclassification by using the \texttt{sslogit} link described in Equation \eqref{eq:sslogit}. Finally we assume that we do not know the sensitivity and specificity but instead assume 95\% probability intervals for each of them, for the sensitivity the 95\% probability interval is $(0.925, 0.975)$, and for the specificity it is $(0.85, 0.95)$. We then follow the approach outlined in the end of Section \ref{sec:response_mc}, that is, approximating the posterior distribution of $p_y$, $\pi(p_y\given \bm{s})$, by summing over a grid of points chosen in the 95\% probability intervals for the sensitivity and specificity weighted by a joint probability distribution for them. This is done through the function \texttt{inla.merge()}, see the supporting information for details. The resulting posterior marginal distributions in Figure \ref{fig:response_mc} show that the models that account for the misclassification better recover the correct disease probability than the model that ignores the misclassification. It also shows that the model that includes uncertainty about the sensitivity and specificity has a wider posterior marginal distribution, reflecting the increased uncertainty in the information we provide the model.

\section{Applications} \label{sec:applications}

\subsection{Birth weight data}
We look at the birth weight data presented in \citet[][Section 1.6.2]{hosmer_lemeshow2000}, an example that was discussed by \citet{buonaccorsi_etal2005} to assess the effects of misclassification error in a binary covariate of a linear regression model. The dataset contains measurement from $n=189$ births, with the birth weight of the child as the response variable $\bm{y}$, the smoking status ($0 = \text{no}$, $1 = \text{yes}$) of the mother ($\bm{x}$) as a binary covariate, and the mother's weight at the time of the last menstrual cycle ($\bm{z}$) as a continuous covariate. For individual $i$, the model is thus given as
\begin{equation}
  y_i = \beta_0 + \beta_x x_i  + \beta_z z_i  + \epsilon_i \ , \quad \epsilon_i \sim\N(0,\sigma^2) \ .
\end{equation}
In this example, the smoking status of the mother is assumed to be subject to misclassification error, while it is assumed that mother's weight does not contain any error. Unfortunately, the misclassification probabilities and the exposure model for $\bm{x}$ are not known, because no validation data were collected in this study. Therefore, we illustrate the error modeling procedure on two hypothetical cases.

\subsubsection*{Birthweight analysis, case 1: Assuming smoking status to be independent of other variables}
\begin{figure}
  \includegraphics[width=\linewidth]{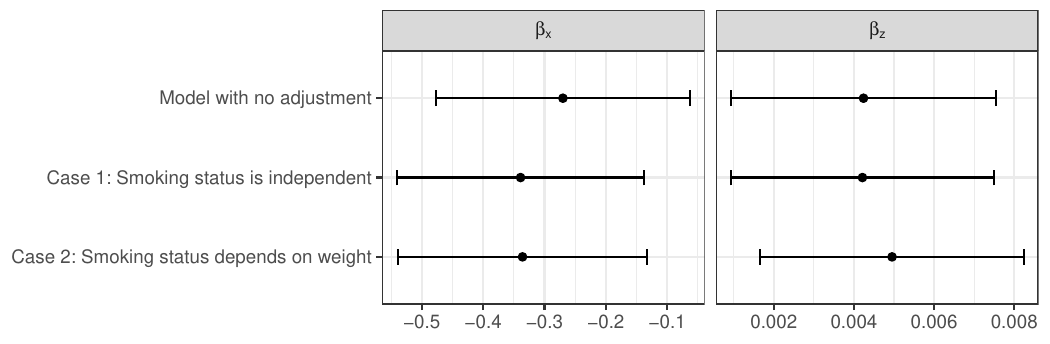}
  \caption{Posterior means and 95\% credible intervals for the naive and error-corrected models for the birth weight example. In case 1, it can be seen that the slope $\beta_x$ for the error-prone, binary variable is corrected with respect to the model without adjustment, whereas the slope $\beta_z$ for the continuous variable $\bm{z}$ is unaffected, because $\bm{x}$ and $\bm{z}$ are assumed independent. In case 2, a bias-correction is now also visible for $\beta_z$.}
  \label{fig:birthweight}
\end{figure}

We assume that non-smokers are less likely to misreport their smoking status than smokers, since pregnant women are not supposed to smoke and might thus try to mask their habit. Therefore, in the absence of better knowledge, we postulate a misclassification matrix given as
\begin{equation} \label{eq:mcmatrix_birthweight}
  \mathsf{M}=
  \left(
  \begin{matrix}
  0.95 & 0.05 \\
  0.2 & 0.8 \\
  \end{matrix}
  \right) \ ,
\end{equation}
thus 5\% of the nonsmokers and 20\% of the smokers report a wrong smoking status. These values are chosen since we expect a very small number of reported smokers to actually be non-smokers, whereas a larger protion of the reported non-smokers are likely actually smokers. A point could also definitely be made that smoking should not be reported as a binary value at all, since a continuous "level" of smoking would be far more accurate. Moreover, we assume that the proportion of smokers in the study was $\P (x_i = 1) = 0.4$, irrespective of the bodyweight of the $i^{\text{th}}$ woman. This number is chosen because in our data, around 39\% have responded that they smoke, and we expect that the correct percentage is somewhat higher.
The exposure model in this case does not depend on $\bm{z}$, which implies that $\bm{x}$ and $\bm{z}$ are assumed independent. Consequently, we do not expect an effect of the error in $\bm{x}$ on $\beta_z$, and the error model should therefore not lead to change in the posterior distribution of $\beta_z$.
Results from \numprint{100000} iterations are shown in Figure \ref{fig:birthweight}.

\subsubsection*{Birthweight analysis, case 2: Assuming association between smoking status and mother's weight}

In the second case, we use the same MC matrix as above, but now assume that the probability that a woman was a smoker depended on bodyweight $\bm{z}$. The respective exposure model, as given in Equation \eqref{eq:exposure}, is assumed known with $\alpha_0=-0.4$ and $\alpha_z=0.02$. The slope is chosen to reflect the assumption that body weight is positively correlated with smoking probability, and the intercept is chosen so that the overall smoking probability somewhat matches the assumed $0.4$\%.
In this model, higher bodyweight increases the probability that a woman was a smoker, which introduces a dependency between $\bm{x}$ and $\bm{z}$. Consequently, we expect that the error model will also have a correction effect on $\beta_z$. Results from \numprint{100000} iterations confirm this (Figure \ref{fig:birthweight}).

\subsection{Cervical cancer and herpes}

\begin{table}
\centering
\caption{HSV data. $\bm{y}$: cervical cancer, $\bm{x}$: accurate test for HSV-2, $\bm{w}$: inaccurate test for HSV-2.}
\label{tab:cervical_cancer}
\begin{tabular}{ l c c c r}
  \toprule
  & $\bm{y}$ & $\bm{x}$ & $\bm{w}$ & Frequency \\
 \midrule
 Validation data &1&0&0&13\\
                 &1&0&1&3\\
                 &1&1&0&5\\
                 &1&1&1&18\\
                 &0&0&0&33\\
                 &0&0&1&11\\
                 &0&1&0&16\\
                 &0&1&1&16\\
 \midrule
 Main study data &1&&0&318\\
                 &1&&1&375\\
                 &0&&0&701\\
                 &0&&1&535\\
 \bottomrule
\end{tabular}
\end{table}

We consider an example mentioned in \citet[][p. 76]{yi2017} and \citet[][Section 8.4 and 9.9]{carroll_etal2006}, originally presented by \citet{carroll_etal1993}. In this study, the connection between cervical cancer and exposure to herpes simplex virus type 2 (HSV-2) is examined. The exposure to HSV-2 is measured through two different procedures, where one is considered to be very accurate (recorded as $\bm{x}$), while the other test is less accurate (recorded as $\bm{w}$). We have measurements using the less accurate procedure for all 2044 patients, whereas only a subset of 115 of them also have so-called validation measurements from the accurate test. The complete data set is given in Table \ref{tab:cervical_cancer}. We also notice that the probability of the HSV-2 exposure being misclassified (that is, the respective entries in $\bm{x}$ and $\bm{w}$ are different) depends greatly on whether or not a patient has cervical cancer. For a patient that has cervical cancer, the probability of falsely detecting exposure to HSV-2 is $\Pr(w_i = 1 \given x_i = 0, y_i = 1) = 0.19$, and the probability of not detecting an exposure is $P(w_i = 0 \given x_i = 1, y_i = 1) = 0.22$. Meanwhile, for a patient that does not have cervical cancer, we instead have $P(w_i = 1 \given x_i = 0, y_i = 0) = 0.25$ and $P(w_i = 0 \given x_i = 1, y_i = 0) = 0.50$. We therefore use an error model with two different misclassification matrices conditional on the cancer status, where the misclassification matrices are estimated from the validation data:
\begin{equation} \label{eq:mcmatrix_bw}
  \widehat{\mathsf{M}}_{1}=
  \left(
  \begin{matrix}
  0.81 & 0.19 \\
  0.22 & 0.78 \\
  \end{matrix}
  \right) \ , \qquad
  \widehat{\mathsf{M}}_{0}=
  \left(
  \begin{matrix}
  0.75 & 0.25 \\
  0.50 & 0.50 \\
  \end{matrix}
  \right) \ .
\end{equation}
We also use a conditional exposure model, based on the sample probabilities in the validation data: $\Pr(x_i = 1\given y_i = 1) = 0.59$, and $\Pr(x_i = 1\given y_i = 0) = 0.42$.

\begin{figure}
  \includegraphics[width=\linewidth]{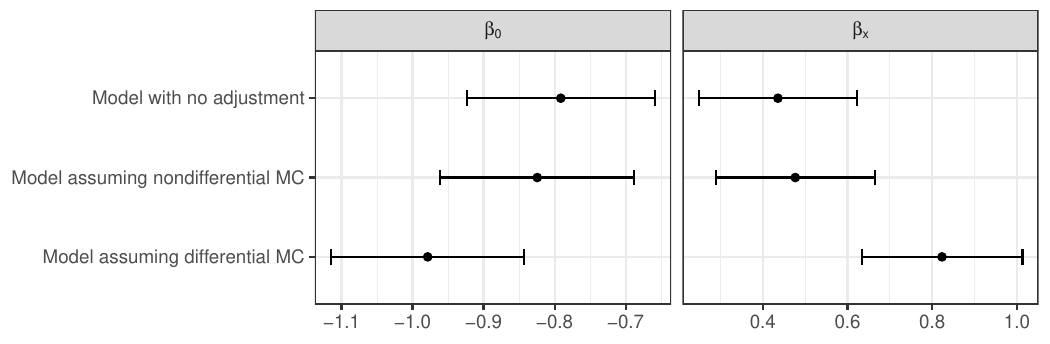}
  \caption{Posterior means and 95\% credible intervals for the naive and error-corrected models for the case-control study on the association between HSV-2 and cervical cancer, assuming either differential or nondifferential misclassification (MC).}
  \label{fig:case_control}
\end{figure}

We fit the model using the INLA-within-importance sampling setup described in Section \ref{sec:is_mc}, using the differential exposure probabitilies and misclassification matrices, and compare it to a model where the exposure probability and misclassification matrix is assumed to be non-differential, as well as a model where no adjustment is done. In both cases we run it for \numprint{100000} iterations. As was also noted by \citet{carroll_etal2006}, we find that the estimates are very different when we assume differential misclassification compared to when we do not (Figure \ref{fig:case_control}). In particular, we can note that the adjustment is much larger for the differential misclassification compared to the modest adjustment we can see when assuming non-differential misclassification.

\section{Discussion} \label{sec:discussion}



In this paper, we looked at methods to account for misclassification and missingness in categorical covariates and responses using INLA. For misclassified or missing categorical covariates, we suggest two different approaches: one based on a combination of importance sampling and INLA, and the other based on an interpretation of the misclassified covariate as a discretized version of a latent Gaussian variable with measurement error. In the former approach, the model for the latent covariate is estimated using importance sampling, and then the model conditioned on that latent covariate is fit in INLA. For the scenario where the observed covariate can be seen as a discretized version of a latent continuous variable with error, we show how this case can be fit using INLA directly, without the need to do any sampling. For response misclassification, we illustrate how the use of a modified link function can account for the misclassification within INLA.

Although the problem of missingness in categorical covariates has been addressed previously \citep[][Section 3.4]{gomezrubio_etal2022}, it has so far not been possible to fit models that account for misclassification in covariates or the response variable using INLA. \citet{gomezrubio_etal2022} outline two methods for imputing missing categorical covariates, one using Bayesian model averaging and another using an INLA-within-MCMC approach. \citep{berild_etal2022} apply their importance sampling approach to a case with a continuous covariate with missingness, but the importance sampling approach has until now not been used in the context of a categorical missing variable or covariate misclassification.


The approaches presented in this paper enable accounting for misclassified categorical variables, and we present conveniently structured code to implement the proposed methods. This makes these methods available for anyone using INLA. However, overall, problems related to misclassified variables remain inconvenient to handle in INLA. This paper presents the results of extensive work on the topic, and proposes the best approaches we were able to find at the current point in time. The main disadvantage with our current implementation of the model adjusting for misclassified covariates is the running time. Although R-INLA is far faster than comparable sampling based methods when used alone, when R-INLA is called within a sampling based method such as importance sampling, it has to be called for every single iteration of the sampling. Because the current implementation of R-INLA creates a large number of temporary files, this means that the running time adds up throughout the iterations and the resulting running time is typically slower than a comparable approach not using INLA. However, fitting the conditional model in R-INLA may still save the researcher time and effort when it comes to implementing the model itself, if the conditional model contains terms that are particularly convenient to model using INLA, such as spatial terms using the SPDE approach or other complex random effects that have good modelling infrastructure in R-INLA. The models will still take a long time to run, but the convenience of the implementation may outweigh the added computational cost.

Part of the reason why the importance sampling within INLA approach requires such a large number of iterations to produce good estimates is that for every iteration, the entire latent covariate is proposed in one block. This means that potentially quite many of the proposed vectors are not particularly good proposals. An alternative could be sampling smaller chunks of the vector at a time, which might prove more efficient.

An alternative to the INLA within importance sampling approach presented in this paper would be to instead use INLA within MCMC. The advantage of this would be that this would allow for also estimating the parameters that the sampling of the corrected covariate values depend upon, namely the coefficients of the exposure model and the misclassification probabilities. In this work, we have assumed them to be known, but this is of course not always the case. However, the disadvantage of using MCMC within INLA is that the procedure is iterative and therefore cannot be run in parallel, like the importance sampling approach can. This means that the method would be even more time-consuming.

The current method to account for response misclassification using a modified link function is still subject to numerical instabilites for smaller sensitivites and specificites, and further work into this topic would be necessary to evaluate if it may be possible to improve the performance in these more critical scenarios. Of course, the cases where the sensitivity or specificity is particularly low are also the cases where it is the most important to account for the misclassification. On the other hand, at least in the context of epidemiology when the misclassified response is a test result (for instance a rapid test for a disease), the sensitivity and specificity of the test are often required to be high in order for the test to be used at all. Additionally, the sensitivity and specificity in these contexts have often been studied extensively, meaning that they can be assumed to be known with quite low uncertainty. In precisely that scenario, our proposed method may work quite well.

The INLA framework is constantly growing in popularity, and accounting for misclassification in both covariates and response variables is important in order to avoid serious biases in inference. Although misclassification error in discrete variables proves to be particularly challenging to handle in INLA, we hope that this paper can serve as a starting point for solving some of these problems, as well as highlighting the challenges that remain.

\section{Computational details}
The code was run on an Intel Xeon E5-2690v4 2.6 GHz CPU using 4 cores. We used R version 4.4, and the functions for the INLA within importance sampling approach is available through the R package \texttt{inlamisclass}, which can be downloaded from GitHub by

\begin{knitrout}\scriptsize
\definecolor{shadecolor}{rgb}{0.969, 0.969, 0.969}\color{fgcolor}\begin{kframe}
\begin{alltt}
\hldef{devtools}\hlopt{::}\hlkwd{install_github}\hldef{(}\hlsng{"emmaSkarstein/inlamisclass"}\hldef{)}
\end{alltt}
\end{kframe}
\end{knitrout}

See the supporting information, available at \url{https://github.com/emmaSkarstein/misclassification_inla}, for specific details on how each example was run.

\section{Author contributions}
E.S., L.B, S.M. and H.R. conceptualized the ideas. E.S. and S.M. did initial analysis and writing on covariate misclassification, while L.B. did initial analysis and writing on response misclassification, and H.R implemented the proposed link in R-INLA. E.S. compiled the manuscript and code. All authors contributed to proof-reading.

\section{Acknowledgements}
The authors would like to thank Sara Martino and Andrea Riebler for helpful discussions.

\bibliography{bibliography}

\end{document}